\documentclass{article}

\usepackage{graphicx}
\usepackage{adjustbox}   
\usepackage{multirow}    
\usepackage{listings}    
\usepackage{longtable}   
\usepackage{array}       
\usepackage{balance}
\usepackage{tabularx} 
\usepackage{booktabs} 
\usepackage{hyperref} 

\newcommand{\Q}[1]{Q$_{#1}$}

\AtBeginDocument{%
  }

\title{Using Large Language Models to Develop Requirements Elicitation Skills}

\author{
  Nelson Lojo\\
  \texttt{nelson.lojo@berkeley.edu} \\
  Univ. of California, Berkeley \\
    CA, USA
  \and
  Rafael González\\
  \texttt{rgonzalez6@us.es} \\
  SCORE Lab, Univ. of Sevilla
  \\ Sevilla, Spain
  \and
  Rohan Philip\\
  \texttt{rohanphilip1108@gmail.com}\\
  Oak Park High School 
  \\ Oak Park, USA
  \and
  José Antonio Parejo\\
  \texttt{japarejo@us.es} \\
  SCORE Lab, Univ. of Sevilla
  \\ Sevilla, Spain
  \and
  Amador Durán Toro\\
  \texttt{amador@us.es}\\
  SCORE Lab, Univ. of Sevilla
  \\ Sevilla, Spain
  \and
  Armando Fox\\
  \texttt{fox@berkeley.edu}\\
  Univ. of California, Berkeley \\
    CA, USA
  \and
  Pablo Fernández \\
  \texttt{pablofm@us.es}\\
  SCORE Lab, Univ. of Sevilla
  \\ Sevilla, Spain
}

\begin{document}
\maketitle
\begin{abstract}


Requirements Elicitation (RE) is a crucial software engineering 
skill that involves interviewing a client and then devising a
software design based on the interview results. Teaching this
inherently experiential skill effectively has high cost, such as
acquiring an industry partner to interview, or training course
staff or other students to play the role of a client. As a
result, a typical instructional approach is to provide students
with transcripts of real or fictitious interviews to analyze,
which exercises the skill of extracting technical requirements
but fails to develop the equally important interview skill
itself.  As an alternative, we propose conditioning a large
language model to play the role of the client during a chat-based
interview.  We perform a between-subjects study ($n=120$) in which students
construct a high-level application design from either an
interactive LLM-backed interview session or an existing interview
transcript describing the same business processes.  We evaluate
our approach using both a qualitative survey and quantitative
observations about participants' work.  We find that both
approaches provide sufficient information for participants to
construct technically sound solutions and require comparable time
on task, but the LLM-based approach is preferred by most
participants. Importantly, we observe that LLM-backed interview
is seen as both more realistic and more engaging, despite the LLM
occasionally providing imprecise or contradictory information. 
These results, combined with the wide accessibility of LLMs,
suggest a new way to practice critical RE skills in a scalable
and realistic manner without the overhead of arranging live
interviews.



    
\end{abstract}

\section*{Keywords}
Computing education, Requirements analysis, Requirements elicitation, Natural language generation, Generative Artificial Intelligence, Large Language Models, Empirical studies, Experimentation

\section{Introduction and Motivation}
\label{sec:intro}

Team skills, including so-called ``soft skills,'' remain an area in which new entrants to the
software engineering workforce are perceived to be underprepared \cite{Dubey2020,Akdur2022}.
One such skill is \emph{requirements elicitation\/} (RE), one component of which involves
interviewing a (usually nontechnical) customer to understand a
business need and propose software-based solutions to support
particular workflows.
In formal higher education, RE is usually taught or practiced by
giving students a transcript of a real or synthetic customer interview
conducted by a team of one or more engineers, and asking students
to specify a technical design that meets the needs expressed in the
interview.  In agile software methodologies, typical specification
artifacts include~\cite{esaas2e} UML class diagrams,
entity-relationship diagrams,  Class--Responsibility--Collaborator
(CRC) cards, and user stories.
Such artifacts would be the inputs to a subsequent team technical
meeting for planning the software architecture.

With this conventional approach to teaching RE,
since students receive a complete interview
transcript, they do not develop interview skills.
To address this drawback, we propose an alternative approach enabled
by large language models (LLMs), in which AI agents take on specific
roles to provide opportunities for students to practice team skills.
We call our approach LEIA, for Learning Enabled by Intelligent
Assistants.  In the case of RE, we
align an LLM to play the role of a
nontechnical customer who knows their business well.  We ask students
to first ``interview'' the customer LEIA (C-LEIA) via a chat-like interface, and then to use
an online tool to produce design artifacts that can be automatically
graded.  By providing differently ``conditioned'' C-LEIAs
corresponding to businesses of varying complexity and customers of
varying temperaments and communication styles, we can give learners
many opportunities to practice RE before interacting with a live
customer.

Our initial user study with computer science
students in a requirements elicitation course at a major research
university focuses on the following research questions:

\begin{enumerate}

  \item[RQ1:] Is the C-LEIA capable of answering relevant questions to provide the necessary and sufficient
    information to produce a valid technical
    solution, while avoiding technical responses that could cue the
      learner?
    
      \item[RQ2:]  Do students using the C-LEIA exercise improve the skill of conducting RE interviews with the goal of constructing technical artifacts? 

      \item[RQ3:] Compared with traditional transcript-based RE exercises, do students find C-LEIA exercises to be an engaging and effective way to practice these skills?
\end{enumerate}

Our results suggest affirmative answers to all these questions.
Section~\ref{sec:related} reviews related work before introducing our
approach and implementation in detail in Section~\ref{sec:leia}.
We describe the design of our user study in Section~\ref{sec:study}
and present observations and results in Section~\ref{sec:results}.
These provide the basis for a discussion on validity threats, limitations, and future work (Section~\ref{sec:limitations}) and
conclusions (Section~\ref{sec:conclusions}).

\section{Related Work}
\label{sec:related}

Customer communication is known to be a major challenge in both
agile~\cite{ElNajar2016,Korkala2009} and non-agile~\cite{Inayat2015}
projects, and the challenge of requirements elicitation 
specifically~\cite{Gizzatullina2019}. 
Eliciting requirements by interviewing can be highly informative and time-efficient, but the quality of information collected
depends strongly on the abilities of the interviewer~\cite{Zowghi2005},
a skill that is difficult to address in conventional classroom
settings:  Daun et al.~\cite{Daun2021} find that “RE is best instructed with experiential learning using collaborative approaches, real stakeholder interactions, or controlled environments simulating realistic experiences, rather than theory-heavy instruction.”
To that end,
\cite{Bano2019, Donati2017} simulate
client-developer interactions with humans, where students conduct
interviews with TAs, are interviewed by TAs, or interview other students,
respectively.

A less instructor-intensive method of practicing RE is to provide a 
student or team of students with the transcript of a real or fictitious 
customer interview, from which the students extract technical requirements. Indeed,~\cite{Gorer2024} uses LLMs to create such transcripts.
But this method does not teach the skill of \emph{how to conduct the
interview\/}, an essential part of the RE process in a professional setting.
To that end, the same authors built~\cite{Gorer2023} an ``interactive robotic
tutor'' to scaffold the interview skill in which
the student is guided through pre-authored structured
questions representing a ``good'' interview, responding to choices
about what to ask next and getting instant feedback for incorrect choices, rather than simulating a truly freeform interview. 

In the pre-LLM era, certain approaches aligned more closely with our work: \cite{Laiq2020} leveraged IBM Watson technologies to simulate a client, creating a more natural conversational flow, while \cite{debnath2020} developed a client interview simulator restricted to a fixed set of predefined responses. Despite these efforts, such methods could lead to awkward or rigid conversations so we believe that the significant recent advances in LLM make this approach worth revisiting.


Finally, LLMs have also been used to ``guide'' interviewers during a RE
interview~\cite{arora2023advancing}; we consider this possibility in
future work, but our present focus is on using LLMs as interviewees
rather than guides or coaches.

\section{C-LEIA: Modeling a Nontechnical Customer}
\label{sec:leia}
\begin{figure}
    \centering
    \includegraphics[width=1\linewidth]{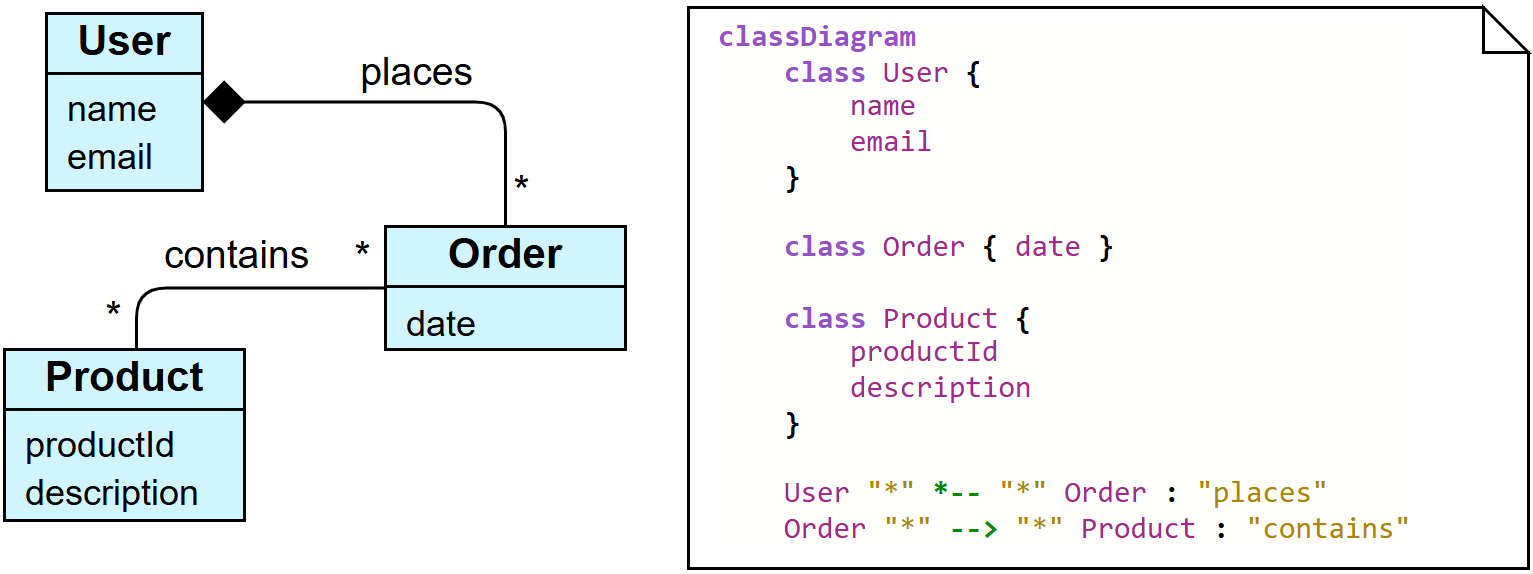}
    \caption{A UML class diagram (left) representing a fragment of an e-commerce site, and its representation (right) in the syntax of Mermaid (mermaid.js.org), a Graphviz-like (graphviz.org) diagram-drawing tool that study participants use to construct their diagrams.}
    \label{fig:uml}
\end{figure}

\emph{Note: If the paper is accepted, we will provide the complete prompt templates, parameter sets, and anonymized participant data in a downloadable replication package. Those details are omitted here due to lack of space.}

We leverage the turn-based nature of interviews to generate dialogue with inference-as-a-service APIs.
Specifically, we align %
an LLM to play the role of a small-business owner (the ``client'') who wishes to have software developed to support one or more business processes.  
Students will then use a chat-like interface to interview the client about these processes and produce a %
UML class diagram (Figure~\ref{fig:uml}) showing a design to model them.
We use the "Assistants" service provided by OpenAI to manage separate conversation threads with the August 6th, 2024 snapshot of GPT~4o parametrized with a default temperature of 1. We use this configuration both for interactive interviews and to generate complete transcripts for our control group, as detailed below.

We focus on two aspects of alignment (i.e. shaping C-LEIA behaviors to reflect its configured intentions), allowing us to describe a customer as a pair of natural language paragraphs.
\emph{Domain alignment} refers to how well the C-LEIA ``knows its
business'' and can respond to questions about the workflows that are
to be supported by the proposed software.  The LLM is guided to integrate information about the reference solution to the specific business modeling problem throughout the conversation, without straying too far or giving too much away.
\emph{Persona alignment} covers various aspects of how a human
customer would respond to questions: is the customer collaborative or
combative?  Are the customer's responses to questions clear and
concise, or are they vague and incomplete, requiring follow-up
questions?
Persona alignment is particularly important with respect to creating a realistic client-interview simulation.
We instruct the C-LEIA to intentionally add filler words, showcase emotions, respond promptly enough to avoid disrupting the pace of the interview, and generally simulate a realistic client conversation through the following guidelines:

\begin{itemize}
    \item The C-LEIA must avoid behavior that reveals its AI nature.
    \item The C-LEIA should have \emph{no technical knowledge of software engineering} and should therefore be unable to respond to technical questions from the interviewer or validate technical choices proposed by the interviewer.
    \item Responses should be natural, concise, and conversational, avoiding lists or overly detailed explanations.
    \item The C-LEIA must not ``guide the conversation'' by providing more information than is requested, responding only to specific inquiries and refraining from using technical terminology or addressing technical concepts.
    \item For vague or broad questions, the C-LEIA should provide general responses and request clarification, as a client without technical expertise might do.
\end{itemize}



In addition to serving as a conversation agent, the C-LEIA was also used to generate complete interview transcripts, such as would be provided to students using the traditional method of teaching RE skills.  We use these for the control group of our study.
For this task, we modified the persona alignment to 
task the C-LEIA with simulating both parties (software engineer and nontechnical client) in the entire interview process, resulting in a full transcript of a simulated interview based on the same solution alignment.
The transcripts were generated using the “Chat Completion” interface rather than the “Assistant” API, maintaining the same model and default temperature of 1.

\section{Study Design}
\label{sec:study}
A common way to teach the RE skills on which we focus is to give students a complete natural language transcript of a real or synthetic interview between an engineer and the client, and ask the student to extract technical requirements from the transcript.
To compare the efficacy of using a LEIA to this traditional approach,
we conducted a between-subjects study.  In the first session, all participants ($n=120$) solve simple design problems using both the traditional transcript-based method and then by interviewing our C-LEIA. In the second session, we randomly assign half the participants to a control group (CG) still using transcripts and the rest to an experimental group (EG) using our C-LEIA, with both groups solving the same significantly more complex design problem.  We then survey both groups regarding their experience with the assignment to shed light on our research questions.

\subsection{Participants}

Participants consisted of three cohorts of students ($n=40$ in each) enrolled in the same Requirements Engineering course at a major European research university.
The study was presented as an optional assignment in the course; 
we offered up to 1 point of extra credit (10\% of the total course grade) towards their course grade for students who chose to participate.  Initially, we announced that half of the bonus would be awarded for participation in the two sessions that conform the experiment, while the remaining half would depend on the quality of their conceptual models, promoting thoughtful and accurate engagement. However, after the experiment was completed, we decided to award the full point to all participants, regardless of performance, in recognition of their commitment to the study.
Students were informed in advance that their anonymized data would be used for publishable research. Students were also informed that they could opt out of the study at any point without penalty; anyone who did so, or who did not wish to participate, could do alternative extra credit activities worth the same amount of credit, ensuring fairness and respecting their autonomy throughout the process.
Students also had access to other opportunities to earn additional credits throughout the course (a total of 25\% additional credits). We believe that this ensured that students did not feel pressured to participate and could engage at their discretion. 

\subsection{Study Design}

\textbf{Session 1,} one hour, all participants: 
    \begin{enumerate}
        \item 10 minutes: introduce Mermaid syntax (Figure~\ref{fig:uml}, right) to ensure proficiency with it in the modeling tasks.
        \item 10 minutes: given an interview transcript implying a simple modeling task requiring 3 entities with 2-3 attributes each, and 2 simple one-to-many and many-to-many associations, use Mermaid to produce a UML class diagram satisfying the design.
        \item 5 minutes: instructor reviews reference solution to discuss potential errors and reinforce good modeling practices.
        \item 25 minutes: use the C-LEIA to interview a synthetic customer regarding a different modeling task requiring 5-6 entities and 4-5 associations, including one composition and reflexive and subclass relationships. This task also included two "open questions," i.e. situations in which more information was needed to complete the model. Use Mermaid to produce a UML class diagram.
        \item 15 minutes: review and evaluate participants' solutions and explicitly address the two modeling open questions.
    \end{enumerate}
\textbf{Session 2:} two hours, with the participants  randomly assigned to equal-sized control group (CG) or experimental group (EG).  This more complex design problem involves 8-9 entities with 2-3 attributes each and 6-7 associations, including two compositions, 1-N and N-N multiplicities, reflexive and subclass relationships, and four ``open questions'' about the modeling problem.
    \begin{enumerate}
        \item CG: Given the transcript (generated by the C-LEIA as described in Section~\ref{sec:leia}) of an interview reflecting the above design, (1) use the Mermaid tool to construct a class diagram, (2) make note of any missing information (open questions) about the design that would be needed to complete the class diagram, (3) identify any information from the transcript that was less essential/lower priority for completing the class diagram.  Participants had up to 105 minutes to turn in a solution, completed or not.
        \item EG: same as CG, but rather than reading a transcript, conduct an interview with the C-LEIA,  which had been aligned to answer questions about the same design problem.
        \item 15 minutes: participants in both groups completed the same survey giving feedback about the assignment, except that a subset of questions on the survey relating specifically to the use of the C-LEIA were administered only to the experimental group.
    \end{enumerate}


In both sessions, after each solution was submitted, the system provided an automated evaluation (developed by the C-LEIA) comparing and identifying the differences between the student’s solution with the pre-configured one and generating a score from 1 to 10. Participants were advised that this preliminary evaluation is not fully accurate or comprehensive and is intended only for information purposes.

\section{Results and Discussion}
\label{sec:results}
\newcommand{\finding}[1]{\textbf{\emph{#1}}}

\begin{figure*}[ht]
    \centering
        \rotatebox{90}{
            \includegraphics[height=\textwidth]
            {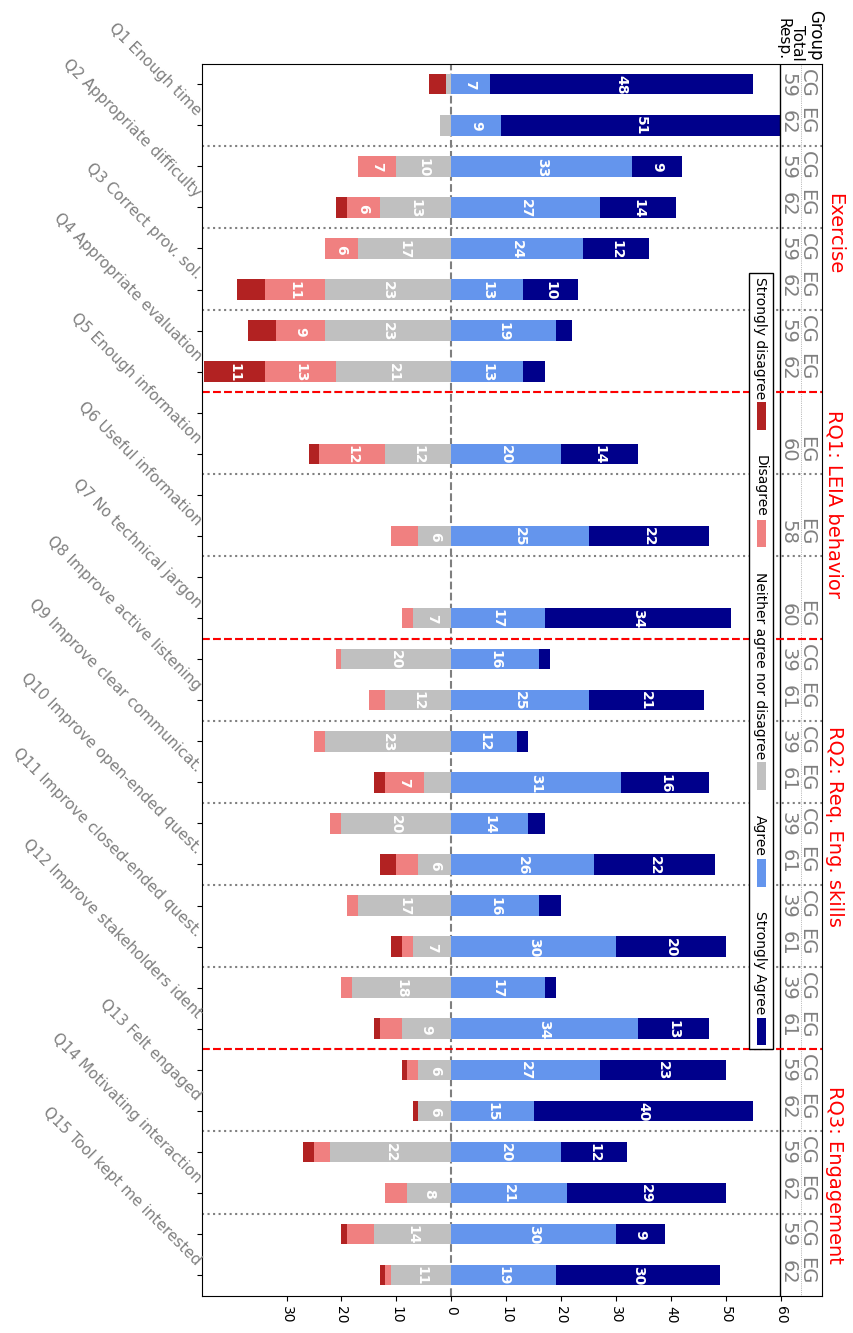}
        }
        \caption{
          Likert-scale responses to survey questions in \autoref{tab:survey_questions}, grouped by Research Question relevance. Since some questions are optional, the total height (total number of responses) is not equal for all questions. 
          In particular, as the text describes, \Q{8-12} had a particularly low response rate in the control group. The text also discusses freeform observations written in the surveys that identify both limitations and possible future work. 
        }
    \label{fig:fullwidth}
\end{figure*}

To address our research questions, we use the survey responses (\autoref{fig:fullwidth}) and various metrics regarding the participants' work products.
Each column of the plot represents one question of the survey (see \autoref{tab:survey_questions}), with the response histograms for both CG and EG for each question.  \Q{5-7} were only asked of the EG.
We summarize each high-level finding in \finding{bold italic}, followed by a discussion of the supporting data.

\begin{table}
    \small
    \centering
    \renewcommand{\arraystretch}{1.25}
    \begin{tabularx}{\columnwidth}{cX}

    \toprule
    
    \Q{id} & Question \\
    
    \midrule


    \Q{1} & The time provided to solve the exercise was sufficient. \\
    \Q{2} & The difficulty of the exercise was appropriate. \\
    \Q{3} & The solution provided by the instructors for evaluating the exercise  made sense. \\
    \Q{4} & The evaluation of my solution (compared with the solution provided by instructors) was appropriate. \\
    
    \midrule
    

    \Q{5}* & (EG only) The C-LEIA provided enough information to solve the design problem. \\
    \Q{6}* & (EG only) The C-LEIA provided useful information. \\ %
    \Q{7}* & (EG only) The C-LEIA avoided technical jargon (regarding software engineering and development) in its responses. \\

    \midrule


    \Q{8}* & The interaction with the C-LEIA helped me to improve as an active listener---understanding both what is said and what is unsaid, identifying the nuances and underlying concerns that may not be immediately apparent. \\
    \Q{9}* & The interaction with the C-LEIA improved my ability to communicate clearly and ask more accurate and precise questions. \\
    \Q{10}* & Interacting with the C-LEIA could help me improve in asking open-ended questions, which are essential for encouraging expansive responses and uncovering missing client needs/requirements. \\
    \Q{11}* & Through interaction with the C-LEIA, I could enhance my ability to ask closed-ended questions, which are crucial for clarifying specific points and ensuring accuracy in understanding client requirements. \\
    \Q{12}* & Interacting with the C-LEIA could help me improve in identifying and managing different stakeholders, each with their own needs, priorities, and influence on the project. \\

    \midrule


    \Q{13} & I felt engaged with the task. \\
    \Q{14} & The interaction with the tool used for gathering requirements (C-LEIA or Transcript) was motivating. \\
    \Q{15} & The tool used to elicit requirements information kept me interested in the activity. \\

    \bottomrule
    
    \end{tabularx}
    
    \caption{Likert scale survey questions (*=optional). The form also solicited free-text comments. \Q{5-7} were only given to the experimental group. \Q{14-15} were answered by the control group with respect to the transcript and by the experimental group with respect to the C-LEIA session.}
    \label{tab:survey_questions}

\end{table}

\subsection{Was the Exercise Valid? (\Q{1-4})}

\finding{Participants understood the tasks, generally found them to be of suitable difficulty, and generally felt they had sufficient time to complete them.\/} A majority agree or strongly agree that the time allocated was sufficient (\Q{1}).
The perception of having sufficient time to complete the exercise was slightly better in the experimental group than the control group, which is interesting for two reasons.  First, interacting with an AI would likely take  more time than working from transcripts: students must think of questions, type them in, and then read the answers to design the model, while students using the transcription only have to read questions and answers.
Second, actual time spent (hours) was comparable between the CG
($\mbox{min}/\mbox{max}/\mu/\sigma = 0.38/1.34/0.86/0.19$)
and the EG
($0.42/1.39/0.85/0.42$).
Although we omit the histograms due to space limitations,
a relatively low Earth Mover's Distance\cite{deza2006} of $0.3708$ suggests the distribution shapes are similar,
though a Sum of Absolute Differences\cite{deza2006} of $26.2246$ suggests distinct patterns in the groups' duration distributions, with wider dispersion in the experimental group.

A majority agreed that the difficulty level was appropriate (\Q{2}).
In both groups, the evaluation of participants' solutions (\Q{4}) appears to have been a source of frustration, particularly due to limitations in the comparison methodology and the simplistic calculation of scores assigned to the models.  We note that in the present study, student scores were not used to validate the research questions, since all of the students' solutions were "above threshold" of correctness for demonstrating an understanding of the material, 
nor were they a factor in the amount of extra credit granted to study participants.
For these reasons, we do not consider these negative responses a threat to validity,
and we felt confident surveying participants about other aspects of the study experience.

\subsection{RQ1: LEIA behavior (\Q{5-7})}

\finding{The C-LEIA provided coherent, sufficient, and useful information for completing the design, mostly without giving away technical aspects of the solution.\/}
EG participants agreed or strongly agreed that it provided sufficient (57\%) and useful (81\%) information; regarding whether the C-LEIA successfully avoided technical jargon responses were even slightly better (85\%).
Overall, responses reveal an overwhelmingly positive experience regarding the behavior of the C-LEIA, with a majority explicitly stating that no erratic, inappropriate, offensive, or incoherent behaviors were observed during the session, and that the C-LEIA facilitated the elicitation process without significant issues. 

\subsection{RQ2: Developing Interview Skills (\Q{8-12})}

\finding{Compared to CG participants, EG participants more strongly agree that the exercise helps them improve key RE skills, including stakeholder identification, question formulation, clear communication, and active listening.}
(P01: "It is more realistic and prepares us better for the future.") 
Significantly, the response rate for these questions was much lower in the CG than in the EG, possibly suggesting that CG participants felt either less motivated or less sufficiently informed to give an opinion on these questions.

\finding{Interviewing is a distinct skill exercised by the C-LEIA as compared to the transcript-based approach.\/} 
While all solutions were substantially correct, the CG's class diagrams tended to include more classes 
($\mbox{min}/\mbox{max}/\mu/\sigma = 6/18/10.61/2.67$)
than the EG's solutions
($4/12/7.68/1.57$).
As well, the total number of class attributes across all classes in the solution was higher for the CG
($\mbox{min}/\mbox{max}/\mu/\sigma = 12/62/28.32/8.83$)
than for the EG
($6/35/15.30/4.97$).
The transcripts given to the CG contain detailed information, whereas the EG must explicitly elicit this information during the interview.  Indeed, in professional practice, it is common to request follow-up interviews as the result of trying to create a design with what is discovered to be incomplete information later.  EG participants are therefore forced to exercise this skill.

\subsection{RQ3: Engagement (\Q{13-15})}
\label{sec:engagement}

\finding{The C-LEIA approach is more consistently able to maintain interest, enhance motivation, and foster overall task engagement.}  Some of this superiority can be attributed to the novelty of the approach, though we tried to mitigate this by giving all participants access to the C-LEIA in Session~1, but some is likely due to the C-LEIA's interactive and adaptive nature. In contrast, the transcript-based approach showed good value but was less consistent in its ability to engage participants, as evidenced by the higher presence of neutral and disagreement responses. Finally, in the freeform comments, EG participants found it highly valuable to ask an unlimited number of questions in their own way to explore requirements (P48: "The AI allows you to ask more questions than you have in the transcript to clarify doubts"), something seen as a key C-LEIA strength that led to deeper engagement with the task.

\finding{An overwhelming majority of participants preferred the C-LEIA over transcripts, with stronger preference among those who used it more.}  The preference for C-LEIA over transcripts was 61\% to 39\% in the CG and 85\% to 15\% in the EG.
The EG’s higher margin of preference is compatible with the conclusion that the C-LEIA not only aligns more closely with skill development objectives, but also  fosters greater engagement and perceived value among participants.

\section{Threats to Validity, Limitations, Future Work}
\label{sec:limitations}

Some of the enthusiasm for using the C-LEIA may be attributable to novelty, though we attempted to mitigate both this enthusiasm and possible nocebo effects~\cite{Kosch2022} (negative expectations about the treatment lead to outcomes that are worse than would have occurred otherwise) by exposing \emph{all} participants to the C-LEIA in Session~1 of the study.
Also, some participants identified weaknesses in our approach or instances of less-than-optimal behavior of the C-LEIA. We summarize a few recurring themes with illustrative quotes from survey responses, using these as a way to suggest future work.  (The language of instruction and of the survey was not English; we used Google Translate to translate both the survey questions and responses for this paper.)

\textbf{Evaluation of student work products.}
Although one motivation for having participants submit a machine-readable rather than hand-drawn class diagram was to allow for the possibility of automated grading, the automatic grading scheme used in this study to compare participants' solutions to the reference solution  was simplistic.  While it was sufficient to ensure that participants' solutions were substantially correct and avoid significant threats to the study's validity, participants felt less positive about this aspect of the study than any other (Q3--Q4). (P15: "The evaluation does not give enough flexibility. There are multiple correct ways to model the exercise.")  We see an exciting opportunity for future work to develop more sophisticated automatic evaluation of these designs, possibly including other design deliverables such as Class-Responsibility-Collaborator (CRC) cards and user stories~\cite{esaas2e}.

\textbf{Inconsistent/contradictory information given.} One recurring issue was the C-LEIA’s occasional tendency to contradict itself, for example, by initially suggesting a relationship between entities and later claiming that the relationship was unnecessary.  Similarly, the C-LEIA occasionally introduced new attributes, properties, or requirements as the session went on, which were not part of the solution. Both kinds of inconsistencies created confusion for participants.
In fact, in practice, this fluctuation of requirements is common with real customers; but for pedagogy, we would like to carefully control when it happens rather than having it happen unexpectedly to a student, as in this case. We anticipate that further LLM alignment will mitigate this problem.

\textbf{Insufficient background information made it challenging to conduct the interview.} 
Starting the interview process without guidance or contextual cues was challenging for some participants, particularly those without prior interview experience. (P42: “Provide some context about the system to be developed, for anyone who is going to use AI. I think it would make the exercise a bit easier and help us ask better questions.”)  In contrast the CG participants only needed to passively read a transcript rather than initiate a conversation.  An immediate mitigation is to include a basic background document that students could read before beginning the interview or reading the transcript.  However, a more interesting future avenue (in our opinion) is a complementary T-LEIA---a "team whisperer" or coach, independent of the C-LEIA, that can give less-experienced students proactive guidance on initiating and conducting the interview by helping to formulate guiding questions, ensuring the discussion stays on track, and so on.
Conversely, to increase the difficulty level (and possibly realism) of the task, we plan to experiment with our persona alignment to simulate "more difficult" customers whose answers lead to a need for followup interviews, either to get more information or clarify ambiguous information from first interview, or customers who are indecisive or unclear in their explanations.


\textbf{Occasional imprecise or less-than-relevant answers to nuanced questions.} While the C-LEIA generally performed well, participants reported that it occasionally struggled to provide precise or relevant answers, especially when questions were very specific or nuanced. Some noted that the AI would sometimes repeat responses or provide ambiguous answers, which could disrupt the flow of the elicitation process. (P88: "If you are too specific, you don’t get enough information. If you are too broad, it feels like you’re trying not to 'break' [the C-LEIA]."; P63: "The main limitation is the precision of the answers.")  

\textbf{Artificial feel of the interaction.}  While the C-LEIA came close to mimicking a real client, many respondents felt that it lacked the subtlety and depth of a human interview, particularly in its inability to convey non-verbal cues or contextual hints (P67: "It doesn’t feel real, although it was close enough"). This artificiality sometimes created a sense of detachment, which limited its effectiveness in fully simulating real-world scenarios. Additionally, the C-LEIA's reliance on carefully structured questions meant that it often failed to provide additional information unless explicitly asked, which could limit the flow of discovery.

\section{Conclusions}
\label{sec:conclusions}
Despite using a prototype C-LEIA whose limitations were noted by study participants, our initial study yielded promising results:
\begin{itemize}
    \item The C-LEIA provides a sufficiently sound and interviewable simulacrum of a nontechnical client representing a small business, providing enough information to produce a viable design without directly disclosing the technical details that underpin the design.
    \item The vast majority of participants clearly preferred using the C-LEIA over static transcripts.
    \item Time on task using the C-LEIA is comparable to that for using transcripts, yet the C-LEIA appears to foster better task engagement and motivation.
    \item Participants using the C-LEIA agree more strongly than those using transcripts that the exercises are helping them improve key RE skills, and in particular, their solutions compared to a control group suggest that interview skills are specifically being exercised.
\end{itemize}

These findings suggest that AI-driven tools hold significant potential for enhancing user engagement in tasks that require sustained attention and active participation like those addressed in the exercise.


The accessibility and convenience of LLMs already provides a promising way to practice requirements elicitation without the logistical complexities of arranging live interviews.
We believe that refining our prototype, and combining it with more sophisticated automatic grading, will
lead to a novel, realistic, and scalable way to teach RE interview skills.

\section*{Acknowledgements}
This research was supported by
Grant \#00005919 from the California Education Learning Lab (calearninglab.org), an initiative of the California Governor’s Office of Planning and Research.
This publication was also partially supported by the R\&D projects PID2021-126227NB-C21, and 
PID2021-126227NB-C22      
funded by MCIN/AEI/10.13039/501100011033/FEDER/UE;
and by the R\&D projects %
TED2021-131023B-C21 and 
TED2021-131023B-C22     
funded by MICIU/AEI/10.13039/501100011033/European Union NextGenerationEU/PRTR.

\newpage
\bibliographystyle{plain}
\bibliography{cleia}

\begin{thebibliography}{10}

\bibitem{Akdur2022}
Deniz Akdur.
\newblock Analysis of software engineering skills gap in the industry.
\newblock {\em ACM Transactions on Computing Education}, 23:1 -- 28, 2022.

\bibitem{arora2023advancing}
Chetan Arora, John Grundy, and Mohamed Abdelrazek.
\newblock Advancing requirements engineering through generative ai: Assessing
  the role of llms, 2023.

\bibitem{Bano2019}
Muneera Bano, Didar Zowghi, Alessio Ferrari, Paola Spoletini, and Beatrice
  Donati.
\newblock Teaching requirements elicitation interviews: an empirical study of
  learning from mistakes.
\newblock {\em Requir. Eng.}, 24(3):259–289, September 2019.

\bibitem{Daun2021}
Marian Daun, Alicia~M. Grubb, and Bastian Tenbergen.
\newblock A survey of instructional approaches in the requirements engineering
  education literature.
\newblock In {\em 2021 IEEE 29th International Requirements Engineering
  Conference (RE)}, pages 257--268, 2021.

\bibitem{debnath2020}
Sourav Debnath and Paola Spoletini.
\newblock Designing a virtual client for requirements elicitation interviews.
\newblock In {\em Requirements Engineering: Foundation for Software Quality:
  26th International Working Conference, REFSQ 2020, Pisa, Italy, March 24--27,
  2020, Proceedings 26}, pages 160--166. Springer, 2020.

\bibitem{deza2006}
Michel-Marie Deza and Elena Deza.
\newblock {\em Dictionary of distances}.
\newblock Elsevier, 2006.

\bibitem{Donati2017}
Beatrice Donati, Alessio Ferrari, Paola Spoletini, and Stefania Gnesi.
\newblock Common mistakes of student analysts in requirements elicitation
  interviews.
\newblock volume 10153, pages 148--164, 02 2017.

\bibitem{Dubey2020}
R.~Dubey and V.~Tiwari.
\newblock Operationalisation of soft skill attributes and determining the
  existing gap in novice ict professionals.
\newblock {\em Int. J. Inf. Manag.}, 50:375--386, 2020.

\bibitem{ElNajar2016}
T.~El-Najar, I.~Ahmad, and M.~Alkandari.
\newblock {Client Communication: A Major Issue in Agile Development}.
\newblock {\em International Journal of Software Engineering and its
  Applications}, 10:113--130, 2016.

\bibitem{esaas2e}
Armando Fox and David Patterson.
\newblock {\em {E}ngineering {S}oftware as a {S}ervice: {A}n {A}gile {A}pproach
  {U}sing {C}loud {C}omputing}.
\newblock Strawberry Canyon LLC, San Francisco, CA, 2nd edition edition, 2022.

\bibitem{Gizzatullina2019}
Ilyuza Gizzatullina.
\newblock {Empirical study of customer communication problem in agile
  requirements engineering}.
\newblock In {\em Proceedings of the 2019 27th ACM Joint Meeting on European
  Software Engineering Conference and Symposium on the Foundations of Software
  Engineering (ESEC/FSE 2019)}, pages 1262--1264. ACM, 2019.

\bibitem{Gorer2023}
Binnur Görer and Fatma~Başak Aydemir.
\newblock {RoboREIT: an Interactive Robotic Tutor with Instructive Feedback
  Component for Requirements Elicitation Interview Training}.
\newblock {\em arXiv preprint arXiv:2304.07538}, 2023.

\bibitem{Gorer2024}
Binnur Görer and Fatma~Başak Aydemir.
\newblock {GPT-Powered Elicitation Interview Script Generator for Requirements
  Engineering Training}.
\newblock In {\em 2024 IEEE 32nd International Requirements Engineering
  Conference (RE)}, volume~33, pages 372--379. IEEE, 2024.

\bibitem{Inayat2015}
I.~Inayat, S.~S. Salim, S.~Marczak, M.~Daneva, and S.~Shamshirband.
\newblock {A systematic literature review on agile requirements engineering
  practices and challenges}.
\newblock {\em Computers in Human Behavior}, 51:915--929, 2015.

\bibitem{Korkala2009}
M.~Korkala, M.~Pikkarainen, and K.~Conboy.
\newblock {Distributed Agile Development: A Case Study of Customer
  Communication Challenges}.
\newblock In P.~Abrahamsson, M.~Marchesi, and F.~Maurer, editors, {\em Agile
  Processes in Software Engineering and Extreme Programming (XP 2009)},
  volume~31 of {\em Lecture Notes in Business Information Processing}, pages
  161--167. Springer, Berlin, Heidelberg, 2009.

\bibitem{Kosch2022}
T.~Kosch, Robin Welsch, L.~Chuang, and Albrecht Schmidt.
\newblock The placebo effect of artificial intelligence in human–computer
  interaction.
\newblock {\em ACM Transactions on Computer-Human Interaction}, 29:1 -- 32,
  2022.

\bibitem{Laiq2020}
Muhammad Laiq and Oscar Dieste.
\newblock Chatbot-based interview simulator: A feasible approach to train
  novice requirements engineers.
\newblock In {\em 2020 10th International Workshop on Requirements Engineering
  Education and Training (REET)}, pages 1--8, 2020.

\bibitem{Zowghi2005}
Didar Zowghi and Chad Coulin.
\newblock {\em Requirements Elicitation: A Survey of Techniques, Approaches,
  and Tools}, pages 19--46.
\newblock Springer Berlin Heidelberg, Berlin, Heidelberg, 2005.

\end{thebibliography}

\balance
\end{document}